# Metamaterial "Multiverse"

Igor I. Smolyaninov

*Department of Electrical and Computer Engineering, University of Maryland, College Park, MD 20742, USA*

**Optical space in metamaterials may be engineered to mimic the landscape of a multidimensional Universe which has regions of different topology and different effective dimensionality. The "metamaterial landscape" may include regions in which one or two spatial dimensions are compactified. Nonlinear optics of metamaterials in these regions mimics either U(1) or SU(2) Kaluza-Klein theories having one or more kinds of effective charges. As a result, novel "photon blockade" nonlinear optical metamaterial devices may be realized. Topology-changing phase transitions in such metamaterials lead to considerable particle creation perceived as flashes of light, thus providing a toy model of birth of an individual physical Universe.**

The newfound freedom of control of the local dielectric permittivity $\varepsilon_{ik}$ and magnetic permeability $\mu_{ik}$ tensors in electromagnetic metamaterials has fueled recent explosion in novel device ideas based on the concept of "electromagnetic space", which is different from the actual physical space, and may have non-trivial topology [1-3]. While current emphasis of research in this field is concentrated in the area of novel electromagnetic devices, linear and nonlinear optics of metamaterials may also have far reaching implications for fundamental physics. For example, optics of metamaterials has a unique capability to realize a table top model of the physical Multiverse [4].

Understanding of our physical world as a tiny fraction of a vast multi-dimensional Multiverse has gained considerable recent attention [4]. String theory, Kaluza-Klein theories [5], and many other higher-dimensional theories, suggest the existence of a landscape of vacua with diverse topologies and physical properties. The landscape generally includes spaces with different numbers of compactified dimensions (Fig.1), with the characteristic size of the compactified dimension being on the order of the Planck length. The symmetries of this compactified internal space define the gauge symmetries and therefore physical laws (types of charges and their interactions) governing a particular region of the Multiverse. Topology-changing transitions between vacua with different numbers of compact dimensions appear to be especially interesting in this model [6]. Such a transition may represent the birth of an individual physical Universe.

Here we demonstrate that using extraordinary waves in anisotropic uniaxial metamaterials, optical models of such space-times as $dS_3 \times S_1$ (3D de Sitter space with one compactified dimension) and $dS_2 \times S_2$ (2D de Sitter space with two compactified dimensions) may be realized. Other non-trivial possibilities, such as the metamaterial models of 4D de Sitter $dS_4$ and anti de Sitter $AdS_4$ spaces [7,8] are also shown in Fig.1. Nonlinear optics of these metamaterial spaces is shown to resemble interaction of charges via gauge fields. Together with recent demonstrations of the metamaterial wormholes [9] and black holes [10-12] which are supposed to connect metamaterial regions having different effective topology, these observations complete the model of the metamaterial Multiverse presented in Fig.1.

Let us start by demonstrating how to produce the $dS_3 \times S_1$ electromagnetic space-time geometry using metamaterials. Spatial geometry of this space-time may be approximated as a product $R_2 \times S_1$ of a 2D plane $R_2$ and a circle $S_1$, as shown in Fig.2(a). Its line element may be written as

$$dl^2 = dx^2 + dy^2 + R^2 d\phi^2 \tag{1}$$

Using the stereographic projection $z=2R\sin\phi/(1+\cos\phi)$, this line element can be re-written as

$$dl^2 = dx^2 + dy^2 + \frac{dz^2}{\left(1+\frac{z^2}{4R^2}\right)^2} \tag{2}$$

Equation (2) indicates that we need a uniaxial anisotropic metamaterial in order to emulate the $R_2 \times S_1$ space. Let us consider a non-dispersive (if the operating bandwidth is sufficiently narrow the dispersion can be neglected) and non-magnetic uniaxial anisotropic metamaterial with dielectric permittivities $\varepsilon_x = \varepsilon_y = \varepsilon_1$ and $\varepsilon_z = \varepsilon_2$. The wave equation in such a material can be written ([13]) as

$$-\frac{\partial^2 \vec{E}}{c^2 \partial t^2} = \vec{\varepsilon}^{-1} \vec{\nabla} \times \vec{\nabla} \times \vec{E} \tag{3}$$

where $\vec{\varepsilon}^{-1} = \vec{\xi}$ is the inverse dielectric permittivity tensor calculated at the center frequency of the signal bandwidth. Any electromagnetic field propagating in this material can be expressed as a sum of the "ordinary" and "extraordinary" contributions, each of these being a sum of an arbitrary number of plane waves polarized in the "ordinary" ($\vec{E}$ perpendicular to the optical axis) and "extraordinary" ($\vec{E}$ parallel to the plane defined by the k–vector of the wave and the optical axis) directions. Let us define a "scalar" extraordinary wave function as $\psi = E_z$ (so that the ordinary portion of the electromagnetic field does not contribute to $\psi$). Equation (3) then yields:

$$\frac{\partial^2 \psi}{c^2 \partial t^2} = \frac{\partial^2 \psi}{\varepsilon_1 \partial z^2} + \frac{1}{\varepsilon_2}\left(\frac{\partial^2 \psi}{\partial x^2} + \frac{\partial^2 \psi}{\partial y^2}\right) \tag{4}$$

Extraordinary field will perceive the optical space as $R_2 \times S_1$ if

$$\varepsilon_2 = n^2 \text{ and } \varepsilon_1 = \frac{n^2}{\left(1 + \frac{z^2}{4R^2}\right)^2} \tag{5}$$

where $n$ is a constant. Similar consideration allows us to construct the $R_1 \times S_2$ optical space shown in Fig.2(b). The line element may be written as

$$dl^2 = R^2(d\theta^2 + \sin^2\theta d\phi^2) + dz^2 \tag{6}$$

Using the stereographic projection, this line element can be re-written as

$$dl^2 = \frac{dx^2 + dy^2}{\left(1 + \frac{\rho^2}{4R^2}\right)^2} + dz^2 \tag{7}$$

where $\rho^2 = x^2 + y^2$. Extraordinary field will perceive the optical space as $R_1 \times S_2$ if

$$\varepsilon_1 = n^2 \text{ and } \varepsilon_2 = \frac{n^2}{\left(1 + \frac{\rho^2}{4R^2}\right)^2} \tag{8}$$

This kind of metamaterial can be formed by multiple single-level quantum wells [14], so that

$$\varepsilon_2 = \varepsilon_\infty - \frac{\omega_p^2}{\omega^2} = \varepsilon_\infty - n(\omega,\rho)\frac{4\pi e^2}{m\omega^2} \quad \text{and} \quad \varepsilon_1 = \varepsilon_s \tag{9}$$

where $n(\omega,\rho)$ is the frequency and coordinate-dependent charge density. In a similar fashion, anisotropic metamaterial described by Eq.(5) can be formed by an array of quantum wires. Thus, optics of metamaterials gives us unique tools to study physics in topologically non-trivial 3D spaces shown in Fig.2. This unique and interesting physics is inaccessible by other means, since these spaces cannot fit into normal Euclidean 3D space.

We are going to consider nonlinear optics of the $R_2 \times S_1$ and $R_1 \times S_2$ metamaterial spaces shown in Fig.2, and demonstrate that it resembles the picture of effective charges interacting with each other via gauge fields. This result is natural, since nonlinear optics of these spaces resembles Kaluza-Klein theories. Somewhat similar picture of mode interaction has been noted previously in the case of cylindrical surface plasmons [15]. Let us start with nonlinear optics of the $R_2 \times S_1$ metamaterial space shown in Fig.2(a), and describe the effective "laws of physics" which arise in this space. These laws are determined by the U(1) symmetry of the internal "compactified" $S_1$ space. The eigenmodes of the extraordinary field can be written as

$$\psi_{kL} = e^{ik\rho} e^{iL\phi} \tag{10}$$

where $L$ is the quantized "angular momentum" number and $k$ is the 2D momentum. The dispersion law of these eigenmodes is

$$\frac{n^2}{c^2}\omega^2 = k^2 + \frac{L^2}{R^2} \tag{11}$$

Therefore, $L=0$ extraordinary photons behave as massless 2D quasiparticles, while $L\neq 0$ photons are massive. Let us demonstrate that in the nonlinear optical interactions of extraordinary photons the "angular momentum" number $L$ behaves as a conserved quantized effective charge.

Nonlinear optical effects deform $\vec{\vec{\varepsilon}}^{-1} = \vec{\vec{\xi}}$ and therefore deform the line element (1). In the weak field approximation corrections to $\vec{\vec{\varepsilon}}^{-1}$ are small. However, corrections to the off-diagonal terms of $\vec{\vec{\varepsilon}}^{-1}$ cannot be neglected. Therefore, the effective metric of the deformed optical space can be written as

$$ds^2 = g_{\alpha\beta}dx^\alpha dx^\beta + 2g_{\alpha 3}dx^\alpha d\phi + g_{33}d\phi^2 \qquad (12)$$

where the Greek indices $\alpha=0, 1, 2$ indicate coordinates of the (almost flat) planar 3D Minkowski space-time: $dx^0=cdt$, $dx^1=dx$, and $dx^2=dy$. This 3D space-time is populated by extraordinary photons (described by "scalar" wave function $\psi$) which are affected by the vector field $g_{\alpha 3}$ (with components $g_{03}=0$, $g_{13}=2\xi_{13}d\phi/dz$, and $g_{23}=2\xi_{23}d\phi/dz$) and the scalar field $g_{33}=R^2$ usually called dilaton. The dilaton field may be assumed constant in the weak field approximation. For a given value of $L$, the wave equation can be written as

$$\hat{D}\psi = \Box\psi - L^2 \frac{1 - g_{\alpha 3}g^{\alpha 3}}{g_{33}}\psi + 2iLg^{\alpha 3}\frac{\partial\psi}{\partial x^\alpha} + iL\frac{\partial g^{\alpha 3}}{\partial x^\alpha}\psi = 0 \qquad (13)$$

where $\Box$ is the covariant three-dimensional d'Alembert operator. Equation (13) looks the same as the Klein-Gordon equation. In 3D space-time it describes a particle of mass

$$m = \frac{\hbar L}{cg_{33}^{1/2}} \qquad (14)$$

which interacts with a vector field $g^{\alpha 3}$ via a quantized charge $L$. The linear portion of $\hat{D}$ in equation (13) describes standard metamaterial optics in which the metamaterial plays the role of a curvilinear background metric. On the other hand, the third order optical nonlinearity of the form

$$g^{(3)}{}_{\alpha 3} \propto \psi \frac{\partial \psi}{\partial \phi} \quad , \text{which means that } \varepsilon^{(3)}{}_{\alpha z} \propto E_z \frac{\partial E_z}{\partial z} \tag{15}$$

leads to Coulomb-like interaction of the effective charges with each other: extraordinary photons having $L \neq 0$ act as sources of the $g^{\alpha 3}$ field, which in turn acts on other "charged" extraordinary photons having $L \neq 0$. The effect of interaction cannot be neglected when the kinetic energy term $\Box \psi$ in eq.(13) becomes comparable with the potential energy terms. For example, this may occur if $k \propto L g^{\alpha 3} = \frac{L g_{\alpha 3}}{g_{33}^{1/2}}$, where $k$ is the momentum component in the $xy$-plane. Therefore, in the limit $k \to 0$ interaction of "effective charges" is important and cannot be neglected.

Nonlinear optics of the $R_1 x S_2$ metamaterial space shown in Fig.2(b) can be considered in a similar fashion. Arising effective "laws of physics" in this case are very interesting because of the SU(2) symmetry of the internal $S_2$ space. Therefore, the extraordinary photons may have two kinds of charges. Motion of these charges is limited to $z$-direction. Since Coulomb-like interaction in the one-dimensional case does not depend on distance, interaction of the "charged" extraordinary photons is very strong: similar to strong interaction of quarks, "charged" photons behave as almost free particles at short distances from each other, while at large distances the potential energy of two "charged" photons grows linearly with distance. It is however limited by metamaterial losses, and by dispersion of $\vec{\varepsilon}^{-1}$. Nevertheless, strong nonlinear optical interactions of extraordinary photons in the $R_1xS_2$ metamaterial space may be used in novel "photon blockade" devices, which are necessary in quantum computing and quantum communication [16].

In addition to non-trivial nonlinear optics, the described toy model of the Metamaterial Multiverse lets us study metric phase transitions [6] in which the topology of the "optical space" changes as a function of temperature or an applied field. A similar topological transition may have given birth to our own Universe. According to some theoretical models [17], during the Inflation our 4D Universe expanded exponentially at the expense of compactification of the extra spatial dimensions. During this process a large number of particles had been created. Metamaterial optics is probably the only other physical system in which a similar process can be observed. Let us consider a topological transition from the $R_3$ metamaterial space to either $R_2xS_1$ or $R_1xS_2$ topology. The number of photons emitted during such metric topology change can be calculated via the dynamical Casimir effect [18,19]. The total energy $E$ of emitted photons from a phase changing volume $V$ depends on the photon dispersion laws $\omega_1(k)$ and $\omega_2(k)$ in the respective phases (see eq.(3) from ref.[18]):

$$\frac{E}{V} = \int \frac{d^3\vec{k}}{(2\pi)^3}\left(\frac{1}{2}\hbar\omega_1(k) - \frac{1}{2}\hbar\omega_2(k)\right) \qquad (16)$$

Equation (16) is valid in the "sudden change" approximation, in which the dispersion law is assumed to change instantaneously. The detailed discussion of the validity of this approximation can be found in ref. [19]. Therefore, the number of photons per frequency interval emitted during the transition can be written as

$$\frac{dN}{Vd\omega} = \frac{1}{2}\left(\frac{dn_1}{d\omega} - \frac{dn_2}{d\omega}\right) \qquad (17)$$

where $dn_i/d\omega$ are the photonic densities of states inside the respective phase. Since the photonic density of states in either $R_2xS_1$ or $R_1xS_2$ phase is much smaller than in the $R_3$ phase, the total number of photons emitted is

$$\frac{dN}{V} \approx \frac{1}{2}\left(\frac{dn}{d\omega}\right)\Delta\omega \qquad (17)$$

where $(dn/d\omega)$ is the usual black body photonic density of states in $R_3$, and $\Delta\omega$ is the frequency interval in which the dielectric tensor of the uniaxial metamaterial satisfies either Eq.(5) or Eq.(8).

In conclusion, we have demonstrated that optics of metamaterials presents us with new opportunities to engineer topologically non-trivial "optical spaces". Nonlinear optics of extraordinary light in these spaces resembles Coulomb interaction of effective charges. Therefore, novel "photon blockade" devices may be engineered. Topology-changing phase transitions in such metamaterials resemble birth of a physical Universe.

**Figure captions**

**Fig.1** According to current understanding of the Multiverse, it is populated by vacua of all possible dimensionalities. A spacelike slice through a multi-dimensional Multiverse is shown. The dark regions represent black holes and black branes. This is a "metamaterial" adaptation of picture presented in ref. [1].

**Fig.2** "Optical space" in anisotropic uniaxial metamaterials may mimic such topologically non-trivial 3D spaces as (a) $R_2 \times S_1$, which is a product of a 2D plane $R_2$ and a circle $S_1$, and (b) $R_1 \times S_2$, which is a product of a 1D line $R_1$ and a sphere $S_2$.

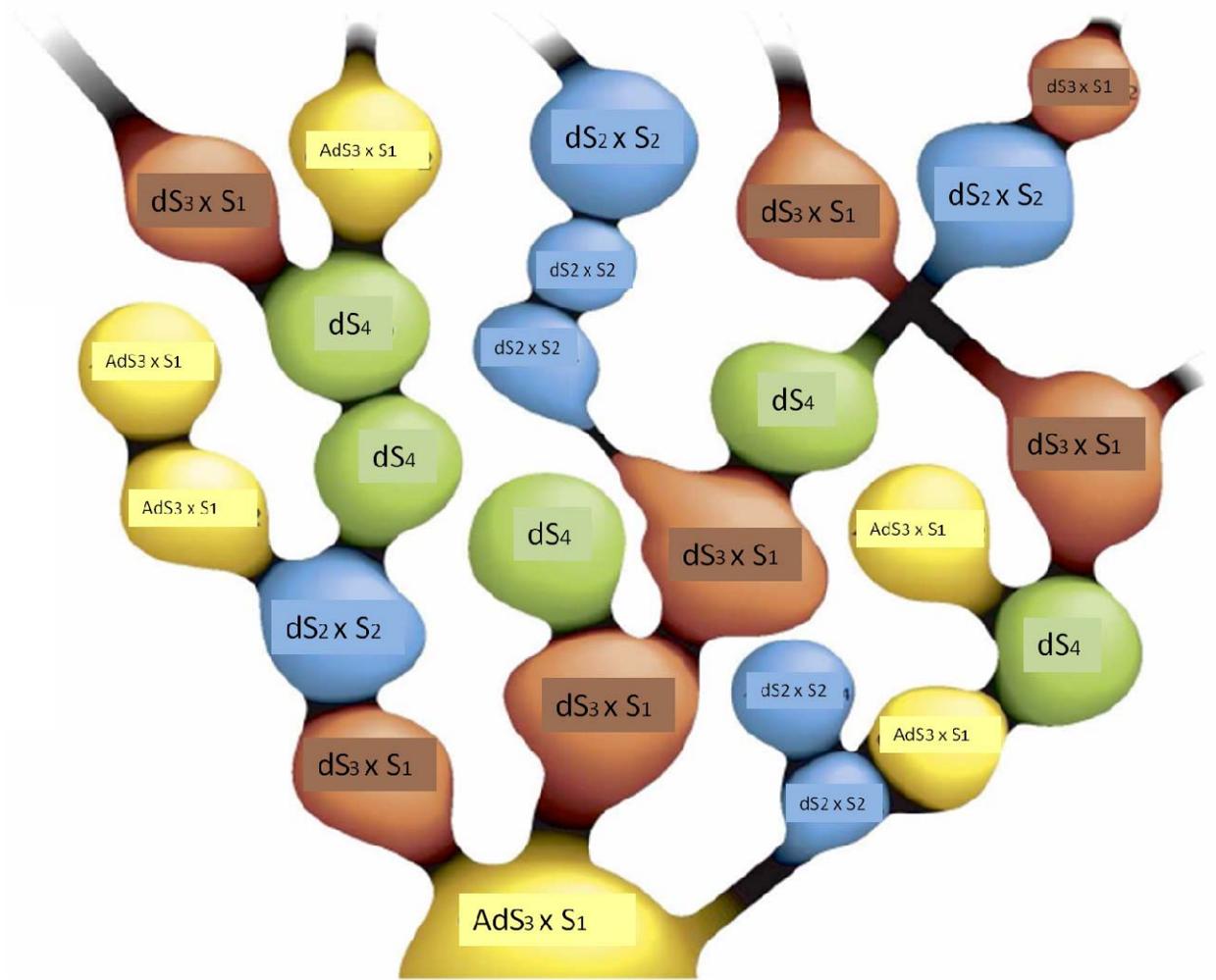

Fig.1

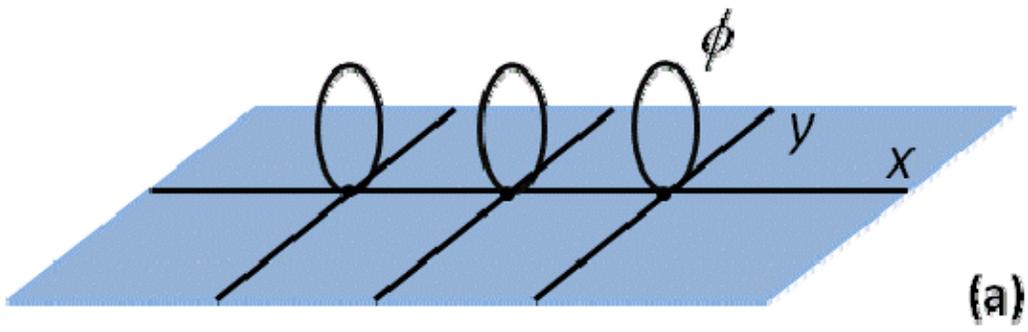

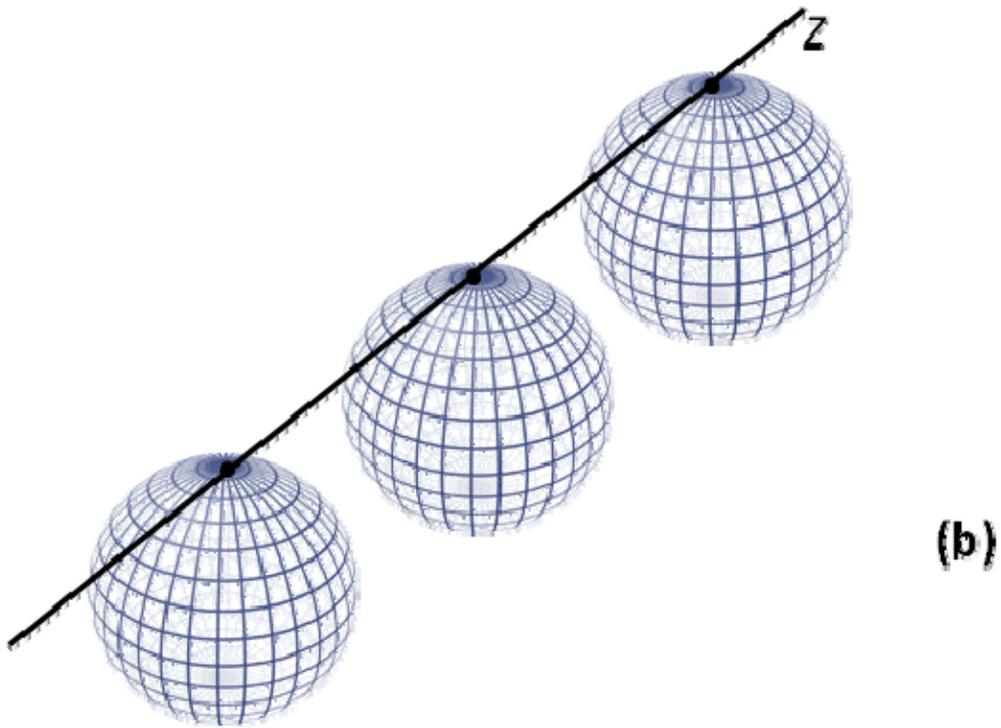

Fig.2